# A model of cortical cognitive function using hierarchical interactions of gating matrices in internal agents coding relational representations

Michael E. Hasselmo

Center for Systems Neuroscience, Department of Psychological and Brain Sciences,

Boston University, 610 Commonwealth Ave., Boston, MA 02215

hasselmo@bu.edu, 617-353-1397

Running head: Cortical function based on internal agents

Keywords: Neocortex, phase coding, predictive coding, spatiotemporal trajectory coding, object transformation, semantic memory

Acknowledgements:  This work supported by the National Institutes of Health, grant numbers R01 MH60013, R01 MH61492 and by the Office of Naval Research MURI N00014-16-1-2832. The author has no conflicts of interest.

**ABSTRACT**

Flexible cognition requires the ability to rapidly detect systematic functions of variables and guide future behavior based on predictions. The model described here proposes a potential framework for patterns of neural activity to detect systematic functions and relations between components of sensory input and apply them in a predictive manner. This model includes multiple internal gating agents that operate within the state space of neural activity, in analogy to external agents behaving in the external environment. The multiple internal gating agents represent patterns of neural activity that detect and gate patterns of matrix connectivity representing the relations between different neural populations. The patterns of gating matrix connectivity represent functions that can be used to predict future components of a series of sensory inputs or the relationship between different features of a static sensory stimulus. The model is applied to the prediction of dynamical trajectories, the internal relationship between features of different sensory stimuli and to the prediction of affine transformations that could be useful for solving cognitive tasks such as the Raven's progressive matrices task.

**INTRODUCTION**

The human cortex appears to play an essential role in a wide range of cognitive functions, and contains many features of neural circuitry that appear to be replicated across multiple regions of cortex (Douglas et al., 1989; Mountcastle, 1997; Rockland, 2010). However, research has not resulted in an accepted, general theory of cortical cognitive function. Computational neuroscience has not fully explored the space of possible structures of neural models of cognitive function. In particular, the features of symbolic processing require the ability to flexibly link cognitive roles in particular rules with specific fillers, which can be arbitrary sensory stimuli or concepts (Fodor and Pylyshyn, 1988; Hummel and Holyoak, 1997). The general capacity for role filler interactions require the ability to identify or imagine a symbolic representation with different location, size or orientation in sensory space, and also the ability to think about a mental concept defined at different positions in the space of internal thought. For example, one can identify or imagine a triangle in visual space, or one can use the same framework to imagine the conceptual analysis of thesis, antithesis and synthesis in mental space. To address the ability to flexibly imagine and manipulate concepts, models of neural function must have the same capacity to flexibly imagine and manipulate activity. These capacities are required for the detection and application of rules in a range of different cognitive tasks (Melrose et al., 2007; Badre et al., 2010; Buschman et al., 2012; Chang et al., 2017; Bhandari and Badre, 2018; Hasselmo and Stern, 2018; Lundqvist et al., 2018).

To obtain this necessary flexibility, this paper proposes a model in which operations depend on highly flexible patterns of neural activity that define connectivity. The model has a larger framework of synaptic connectivity, but the current influence of connections (referred to as gating matrices) depends upon the state of neural activity within individual components of the model. The changing patterns of neural activity in the model are referred to as internal agents. The term agent is used based on the capacity of separate activity patterns representing individual internal agents to independently explore and build models of the internal state space of the network. The term internal is used to contrast with agents that explore the external environment and build behavioral models in reinforcement learning theory. These internal agents correspond to activity patterns that can guide different internal operations on internal representation. Based on the large neural populations available in cortical structures, cortical circuits would have numerous internal agents operating simultaneously. The internal agents can represent a current index of neural activity being tested by the agent, and the current transition between indices. Thus, instead of mapping input vectors (either static vectors or sequences) to specific output vectors (either static vectors or sequences), the network uses multiple internal agents to explore the internal state space and sensory input space, to discover gating matrix functions that account for patterns of activity. Thus, a given gating connectivity matrix can shift based on activity into many different configurations, as in previous models (Badre and Frank, 2012; Kriete et al., 2013; Hasselmo and Stern, 2018). The influence of activity on connectivity could be mediated by different neural mechanisms, such as synapses on dendritic trees that mediate multiplicative interactions of adjacent inputs, allowing activity at one synapse to show multiplicative interactions with adjacent synapses. These multiplicative interactions mediate updating of the indices of internal agents at one level and the influence of higher level internal agents on the internal matrix element values of lower level agents.  The representation of indices could involve phase coding of sensory locations and indices.

The model has the following characteristics.

1. Network gating matrix connectivity depends upon the current pattern of activity within internal agents.

2. Higher level internal agents update the internal gating matrices of lower level agents.

3. Activity is defined in terms of indices defining locations in sensory space or in the neural space coding cognitive operations.

4. Activity is combined and updated based on internal agents that explore and update indices in sensory or neural space.

5. Relations between indices can be discovered or defined by functions generated by internal gating matrices.

This model relates in a general manner to previous work on modeling the representation of rules and concepts in neocortical circuits.  In particular, this extends a previous model using neural activity to gate matrices of connectivity between different structures (Hasselmo and Stern, 2018). This work relates to previous models of the flexible gating between different cortical working memory buffers (O'Reilly and Frank, 2006; Kriete et al., 2013) consistent with previous theories of flexible routing of information for rule learning in prefrontal cortex (Miller and Cohen, 2001; Wallis et al., 2001; Wallis and Miller, 2003). These relate to the general area of modeling the formation of flexible relational representations in neocortex, which was previously explored in neural blackboard models of cognition (van der Velde and de Kamps, 2006, 2015) or models using synchrony of oscillations (Hummel and Biederman, 1992; von der Malsburg, 1995; Siegel et al., 2009), or compression operators to link codes (Eliasmith et al., 2012; Eliasmith, 2013). The model presented in this paper demonstrates how the concept of flexible gating could be implemented by circuits within neocortex to use internal functions to represent visual images as well as behavioral trajectories.  Previous models used grid cell representations to code behavioral spatiotemporal trajectories to store episodic memories. The individual locations along the spatiotemporal trajectory could then be associated with specific objects or events coded on these trajectories (Hasselmo, 2009; Hasselmo et al., 2010; Hasselmo, 2012). Similarly, the relationship between the position of visual landmarks was used to code the current location of the agent in the environment (Byrne et al., 2007; Raudies and Hasselmo, 2015; Raudies et al., 2016). As proposed in an earlier paper (Hasselmo et al., 2010), the framework presented here moves beyond the coding of an episodic memory to the coding of semantic representations of the environment by the relations between different component features, such as the relation between a set of visual cues at different locations and a reward location that might be used to code a specific familiar environment or cognitive task independent of individual episodic memories.  The use of gating agents also builds on previous models using interacting populations of neurons to gate selection of motor actions (Hasselmo, 2005; Koene and Hasselmo, 2005) or to make context-dependent choices in behavioral tasks (Hasselmo and Stern, 2018). This also relates to models using actions to load information into working memory or episodic memory for solving behavioral tasks (Hasselmo, 2005; Hasselmo and Eichenbaum, 2005; Zilli and Hasselmo, 2008), as well as models in which internal models of behavioral trajectories are used to plan future actions (Erdem and Hasselmo, 2012; Kubie and Fenton, 2012; Erdem and Hasselmo, 2014; Bush et al., 2015; Erdem et al., 2015).

One fundamental feature of this model is the direct influence of gating neurons on the synaptic spread of activity between other neurons, which could arise via different possible mechanisms. This gating could involve synapses on the dendritic tree of pyramidal cells that interact via voltage-dependent NMDA conductances, such that a conjunction of nearby synaptic inputs is necessary to generate an output giving a multiplicative function (Mel, 1993; Poirazi et al., 2003; Nezis and van Rossum, 2011). This model is supported by the fact that neurons receive multiple different functional synaptic inputs, but commonly only respond to a subset of these inputs (Chen et al., 2011; Chen et al., 2013). Alternately, a multiplicative interaction of activity can be mediated by network interactions (Nezis and van Rossum, 2011), or it could involve axo-axonic inhibitory interneurons that can directly regulate spiking output (Klausberger et al., 2003; Cutsuridis and Hasselmo, 2012) or interactions between the oscillatory dynamics of different populations (Sherfey et al., 2018).

This model does not use the modification of synaptic weights by gradient descent commonly used in deep learning models (LeCun et al., 2015). In contrast, the current model focuses on rapid gating of synaptic connections based on interactions of activity patterns such as multiplicative interactions of adjacent NMDA synapses. The use of internal gating agents in the model focuses on formation of independent predictive function matrices that model the sensory or conceptual space.

**METHODS**

**Model structure**

The overall structure of the model elements include different levels that could be considered analogous to different cortical regions which function at different hierarchical levels of information processing.  As shown in Figure 1, these regions include a sensory input map, and internal gating agents that function across different hierarchical levels. An internal gating agent refers to an individual, independent functional unit defined by patterns of neural activity, that can search and predict in a given state space, and function across multiple different levels. The searching involves gating to sample specific indices and input values in sensory space or neural circuits. The prediction involves building functions that can approximate and predict patterns of indices and values. The state space of a given unit depends on its level within the processing hierarchy, such that low-level agents search and predict among sensory input stimuli, and higher-level components of agents can search and predict matrix elements in the matrices or vectors of lower level agents.

**Sensory input map**

The sensory input map is a two-dimensional array with a pattern of sensory input similar to the retinotopic map in primary visual cortex associated with a specific visual stimulus, or the map in somatosensory cortex.  The sensory map is defined as a two dimensional array $M_{x,y}$ with indices for position x and y.  Grayscale images consist of continuous values ranging between 0 and 1 for each location indexed by x,y.  Internal gating agents can sample different indices of positions x and y and perform operations on the grayscale values, to allow the formation of functions approximating the sensory pattern.

The function of the model is implemented with the representation of sensory input by arrays of column vectors containing elements coding the indices of location and the corresponding sensory values.  In the initial examples presented here, the individual column vectors of sensory input arrays consist of 3-element vectors with an index for x and y location and a sensory value between 0 and 1. For example, the sensory input in Figure 2 is black (value 1), but the sensory input in Figure 4 is dark gray (0.99).  In the examples presented here, gating matrices are computed to represent the relations between the indices of the components of the sensory input.  The model selects subsets of locations and values to create n by n arrays of sensory input that are operated on by n by n gating matrices that code the relations between the values at different sensory input locations. In the examples presented here, the sensory input is sampled as individual single column vectors coding one location of input, or as 3 x 3 arrays of column vectors coding three locations of input. The next level of the model computes relations between these sensory input arrays using 3 x 3 gating matrices.

**Internal agents.**

The next region or level of the model contains activity patterns that form internal gating agents that explore and form approximations of the properties of the sensory input map, in the form of relations between small arrays of sensory input vectors. Each internal gating agent could be analogous to a cortical column, but the details of this relationship are not elaborated here. The activity within agents in the first level regulate the gating of interactions with the sensory map to sample sensory input at the indices for specific sensory locations (analogous to retinotopic position in primary visual cortex or somatosensory position in primary somatosensory cortex, or to allocentric spatial location in entorhinal cortex or hippocampus). The pattern of activity within an internal agent consists of vectors maintaining information and activity patterns that act as gating matrices.  Thus, the gating matrices described here usually correspond to patterns of gating activity rather than synaptic weight matrices (though they could in some implementations).

The circuitry of an internal agent can vary, but these internal agents can contain different functional components listed here and shown in Figure 1. An internal gating agent can contain activity representing subsets of the following: 1. the current agent attentional focus position vector, 2. a test location vector for next position being sampled for location and value, 3. a prediction or relation matrix, which in the examples presented here link sensory arrays representing location and value to other arrays of location and value, 4. a search matrix for guiding the next test location, and 5. a comparison process for evaluating the sensory input at the next test location relative to the sensory input of the current agent sample position. The sampling of sensory input could take many forms, but here the description will focus on detecting and following simple high contrast trajectories or edges in a visual input.

These different component processes of the agents require different time scales. The actual behavior of an animal in an environment would be described here as real time. The current focus of attention within a representation would move according to attention time. The sampling of the next focus of attention could involve a faster scanning process on the scale of sampling time.

Consistent with the expansion of neuron number in sensory association cortices of the human brain relative to primary sensory regions, the first region or level containing internal agents could potentially contain a much larger number of neurons than the sensory map.

**Mechanism of prediction**

The neural activity within cortical circuits are here modeled as internal gating agents with evolving patterns of neural activity.  The dynamics of each internal agent at this level generates activity vectors $\boldsymbol{a}$ that represent the current attended index of an anatomical/spatial location that is the focus of attention at that point in attention time in the sensory input map.

The focus of attention of the internal agent moves across the sensory input map either according to a predicted relation between the current focus and the next test location, or on the basis of a search process. At each time step of attention time, all the individual internal agents within a region generate a prediction matrix about the index of the next location with a similar value.  In the first example shown here, when the value of sensory input at the current attentional focus index location matches the sensory input value at the next sample index location , then the internal agent predicts the relationship between the indices at sequential positions by computing the gating matrix that will lead from the array of three column vectors coding sensory location indices X(t) at the current focus for focus time step t to an array of column vectors coding sensory location indices X(t+1) at the next focus time step t+1. (See Figure 1). This next focus might be determined after a number of search steps in sampling time. The gating matrices in this model are represented by the letter $\boldsymbol{G}$. The prediction gating matrix $\boldsymbol{G}$ between these sequential arrays can be determined by network computation as follows using the next focus array and the inverse of the current focus array:

(1)      $\boldsymbol{G = X(t + 1)X(t)^{-1}}.$

Once a prediction matrix is generated, it can be tested on subsequent time steps by using the prediction matrix based on the current input X(t) to predict the array ending with the next sensory input index X(t+1). This process can be repeated for multiple steps k to predict a full trajectory of future locations, by raising the prediction gating matrix to different powers k in the following equation as follows:

(2)      $\boldsymbol{X(t + k) = G^k X(t)}$

Examples of the prediction function are shown in the first example in the results section. In later transitions through the same sensory indices, the output of location and value from the prediction function could be compared with the actual locations and values of the sensory input and the prediction function can be maintained on the basis of its success in prediction. Here the comparison of the predicted and actual locations are shown in the first example in the results section. In future work, the accuracy of prediction could be used to select internal gating agents, and allow evolution of the structure of gating agents.

**Mechanism for search or sampling**

If no prediction matrix has been created by an internal agent, or if the prediction matrix does not generate an accurate prediction, then the agent generates a search or sampling process within a given attention time step. Within this attention time step a set of sampling time steps can allow a sample gating matrix of the agent to generate activity to sample the index of the sensory input location being tested. The selection of an input index location by the sample gating matrix allows transfer of the sensory input value (e.g. gray scale in the examples) from that index location to the agent sampling this location. Or alternately, the selection of a particular value could allow the transfer of the corresponding location indices.  Thus, rather than a full spread of forward activity along a feedforward connectivity matrix, the network uses a pattern of bidirectional connectivity that allows the activity of internal agents to explore and model the sensory input space. For a single agent, the current sampling/testing location of the agent is defined by a 2 or 3 element vector $a$.  The search strategy for a single agent is defined by the activity pattern of a matrix $S$ called the sampling matrix and a sampling vector $s$. On each sample time step m, the test location vector $o$ being tested by a single agent is described by:

(3)      $o = a + S^m s$,

where the initial test location relative to current location is defined by $s = [x \quad v_x \quad y \quad v_y]$. This represents activity spreading different numbers of cycles m through a recurrent matrix from the initial activity. This describes a search pattern that moves progressively through the indexed space of the sensory map, allowing testing of the sensory value for different indexed locations. An example of the search process is shown in the 2nd example in the results section. Internal gating agents could function in many different ways to create different test search patterns, and the type of search pattern could be modified to provide the information necessary to solve specific types of cognitive tasks.  In some cases, the prediction matrix generated from the inverse in equation 1 may be poorly conditioned. In this case, the search pattern might have to use different combinations of sensory values to generate the prediction matrix

The sampling process also involves a comparison process. In the 2nd example presented here, the comparison process involves sampling across different locations and comparing the sensory input values at sample locations with the range around a sensory value in memory that guides the sampling. When the values do not match sufficiently, then the current sampling time continues and the cycles m increase. When the values do match within a range, the sample time stops at cycle m, and the matching sample value becomes the test value used for generation of a new prediction or relation as described in the previous section. After generation of a prediction matrix, the new test location becomes the current attentional focus, and the new sampling cycle starts from this new focus location, as shown in the 2$^{nd}$ example in the Results section.

**Higher level components of internal agents.**

Higher level components of internal agents correspond to further levels of association cortex with cortical columns with the same dynamics of internal agents, but with different functional roles acting on lower levels of agents. The different levels could act within a single internal agent, or could involve interactions of an independent higher level agent with multiple different lower level agents. Thus, an internal agent can be either a localized pattern of activity in one region, or an interaction of activity patterns across regions. Rather than exploring the indices of the sensory input space, these higher levels explore the state space made up of the activity vectors of components of the internal agent at the next lower level, or combinations of component elements of the gating matrices at the lower level.  Thus, the higher level components of agents seek to model the internal consistencies of gating matrices at the lower levels of either the same agent or other agents.

For example, the transition points between different objects or different trajectory components involve a change in the internal prediction matrix of an internal agent.  A couple of examples are presented here. For example, in the 1$^{st}$ example of the trajectory of a ball (Fig. 3), the first level representation is the array of vectors representing locations along the trajectory, the second level representation is the gating matrix linking sequential positions (computing velocity), a third level representation is the gating matrix linking sequential velocities (computing acceleration). The next higher level generates the dynamical prediction matrix. At an even higher level, if a trajectory of a ball bounces against a barrier, the initial condition vector changes. This can be coded by a higher levels of the internal agent being activated by the convergence of ball position with barrier position, and updating the initial condition vector to reverse the current velocity, thereby linking one trajectory segment to the next trajectory segment.

As another example in the Results section, the exploration of sides of shapes such as rectangles involve multiple independent agents that move in different directions across the shape and along the edges to become localized in the corners. The position of these different agents can then provide a useful set of parameters to be acted on by higher level agents described in a further section of the Results.

The subsequent section of the Results section involves interaction of higher level agents with the corner representations that could be formed by the lower level agents in the previous part of the results. The higher level agents start with the position of corners, and then perform further levels of representation. When learning the shapes and sequence of shapes in a Raven's progressive matrices task, the transformation between corners of the stimuli or shapes might involve an affine rotation transformation, while the transition between different shapes might involve an affine transformation such as translation and a change in rotation or size. This can be represented as a transformation of the transformations of the corners. In this framework, a first level of an agent might code the corners of a shape as arrays of three vectors representing locations in sensory space. As described below, the second level of an internal agent could code the relation between the corner array and a reference corner array. A third level of an internal agent could then code the relationships between the second level agents, thereby coding the transition between different corners. A higher level of the agent could then code the relationship between different shapes, to code the translation between different shapes and the transformation of the shapes such as a change in rotation or size or stretch. These higher level matrices can be used to learn transformations for existing relations between corners in shapes and extrapolate them to further shapes as described in the Results section.

At all levels, the internal gating matrices could also undergo branching effects, to allow a single trajectory or feature to simultaneously branch into multiple other trajectories or features. This would allow more rapid prediction of sensory features than a purely sequential process. For example, a single higher level of the agent could predict multiple elements in a lower level of the same agent or multiple agents using gating matrix tranformations simultaneously rather than sequentially.

Because the internal gating matrices are patterns of activity, they could also be transferred or copied across different internal agents by higher level agents that copy the matrices to different agents via a higher level gating matrix. This would allow a function or trajectory or rule that was learned in one modality to be rapidly transferred to another modality via transferring the necessary elements of the functional gating matrices that consist of patterns of activity.

### Internal agent generators.

The internal agents could be coded genetically and generated locally, but could also reproduce their activity patterns to replicate specific functional modeling capabilities. That is, there could be internal agents that are specialized for generating the functional properties of specific internal agents.  Because the features of the internal agents are defined by patterns of activity, they could be replicated and propagated by transferring the relevant activity pattern of the starting source agent.  The internal agent generators could be created by longer-term processes that involve either genetic coding or learning of particular patterns of long-term synaptic connectivity that can generate activity patterns representing internal agents at specific time points or cortical indices. Thus, internal agents could be generated by internal agent generators that use modifiable synapses to store patterns that can generate the activity matrices and vectors of internal agents at multiple different levels or times.

### RESULTS

This section describes some applications of the internal agent formulation to demonstrate how this could provide a framework to form cortical representations of functions relevant to cognitive processing. This section focuses on three examples. The first example involves learning and predicting the physics of projectiles or other dynamical trajectories.  The second example involves sampling the sensory values of visual images to detect edges and corners of shapes in the image.  The third example takes the output of the $2^{nd}$ example for coding corners, and uses it for coding visual images representing components of a Raven's progressive matrices task, and then making predictions or extrapolations of the affine transformations that are a component of Raven's progressive matrices.

**Predicting spatiotemporal trajectories**

Many behaviors such as catching a ball require understanding the physics of a sensory situation. For example, Figure 2A illustrates the example of a ball that has been thrown in the air with a certain velocity in the x and y direction that then follows a parabolic trajectory due to the force of gravity exerting acceleration in the y direction. In order to guide behavior based on an object such as a ball moving under the influence of initial velocity and gravitational acceleration, neural circuits must be able to form an internal model of the dynamics of the trajectory of the object. On a more general level, the capacity to match the dynamics of a system could be used for internal representations of a range of different curves in space and time. The following description addresses how a neural circuit could potentially code functional dynamics in a manner similar to physical trajectories. This could involve a specific internal agent that responds to sequential detections of the object and computes the velocity and acceleration.

An internal agent can compute the velocity and acceleration of the trajectory based on the sequential position of a minimum of three features on an object, each coded by column vectors such as $a_1(t)=(x_1(t); y_1(t); 1)$. The overall structure of such an agent is shown in Figure 3A, and the sequential computation described here is shown in Figure 3B. At each time point, the three feature column vectors a1, a2, a3 at three different x,y positions can be coded in an array matrix $X(t) = \begin{matrix} x_1(t) & x_2(t) & x_3(t) \\ y_1(t) & y_2(t) & y_3(t) \\ 1 & 1 & 1 \end{matrix}$. The bottom row allows the computation of affine transformations. At the next time point, the array has been updated to matrix $X(t + 1)$. (Note that the order of the features can be important as discussed below, but if they are spatially very close to each other relative to the distance between then their relative order matters less).

As shown in equation 1, the prediction matrix describing the translation between consecutive time points can be computed by the computing X(t+1) times the inverse of X(t) as follows: $_{r=2}G(t + 1) = X(t + 1)X(t)^{-1}$, where r=2 indicates the level of the representation. This yields a second level prediction gating matrix functionally associated with the same internal agent that includes the affine translation of the object between the two time points. This matrix G(t+1) includes the translation terms $v_x$ and $v_y$ in the third column of the 1$^{st}$ and 2$^{nd}$ rows that correspond to velocity.

Taking the next level matrix operation between the second level gating matrices yields a third level gating matrix as $_{r=3}G(t+1) = {}_{r=2}G(t+1)({}_{r=2}G(t)^{-1})$, where r=3 indicates the level. This third level matrix is functionally associated with the same internal agent and contains the affine translation terms corresponding to acceleration $a_x$ and $a_y$ dimension in the third column of the $1^{st}$ and $2^{nd}$ rows.

The next step requires taking the position, velocity and acceleration terms from the above matrices at different time steps and using them to construct a new array for use in generating a prediction matrix (equation 1). This construction of the array could be performed by an index transformation matrix that translates the relevant portions of the above gating matrices into two separate 3-dimensional arrays (for x and y) that code the position, velocity and acceleration at each time point. The array at time t would be:

$${}^{va}X(t) = \begin{matrix} x(t-2) & x(t-1) & x(t) \\ v_x(t-2) & v_x(t-1) & v_x(t) \\ a_x(t-2) & a_x(t-1) & a_x(t) \end{matrix}$$ , and the array at time t+1 would be:

$${}^{va}X(t) = \begin{matrix} x(t-1) & x(t) & x(t+1) \\ v_x(t-1) & v_x(t) & v_x(t+1) \\ a_x(t-1) & a_x(t) & a_x(t+1) \end{matrix}$$ . Corresponding arrays would be created for the y

dimension. The example here uses cartesian coordinates, but could be applied in other coordinate systems such as the coordinates at 60 degree angles used by grid cells.

The construction of these arrays requires complex index transformations that could be coded by tensors transforming elements from one matrix to another matrix, with buffering of information from previous time points. For example, an index transformation tensor would have the value 1 at the index that translates the acceleration $a_x(t)$ from the third level gating matrix row 2 column 3 to array X(t) position row 3 column 3. The complex index transformation tensor could be genetically pre-programmed in cortical circuits or could be obtained during development by an activity-based selection process of transformation matrices in a large number of internal agents with multiple different connection patterns.

An inverse operation performed on these two array matrices will yield the prediction gating matrix that describes the dynamics of the object in each dimension in the absence of a collision as $_{r=4}G(t+1) =$ $_{r=3}^{vq}X(t+1)(_{r=3}^{vq}X(t)^{-1})$. This prediction gating matrix can predict the remaining components of the trajectory when given an array matrix from a specific position on the trajectory.

Figure 2 illustrates the effectiveness of this framework for determining the dynamical equation determining physical trajectories such as parabolas, sinewaves, circles and spirals.  For example, as shown in Figure 2A, the dynamical prediction matrix for a parabola can be computed after 6 steps (open circles). This dynamical prediction matrix can then be given an array matrix from the trajectory and used to accurately predict the rest of the parabola trajectory. The asterisks within each circle show the predicted trajectory based on the dynamic prediction gating matrix and variables determined at step 6. That is, in the example in Figure 2A, the prediction starts with the array matrix from steps 3-6. No further input is given to the model, but it accurately traces the full trajectory with the asterisks inside the remaining circles. As an example, the y dimension of the parabola is computed by the dynamic prediction matrix: $\begin{array}{ccc} 1 & 1 & 0 \\ 0 & 1 & -.2 \\ 0 & 0 & 1 \end{array}$.  Other parts of Figure 2 show this same predictive function based on 6 steps (open circles) with asterisks effectively predicting the trajectory of sinewaves, circles, spirals and other shapes. As a further example, in Figure 2B, the sinewave has the y dimension prediction matrix: $\begin{array}{ccc} 1 & 1 & 0 \\ -.01 & .99 & 0 \\ 0 & 0 & 1 \end{array}$, and the circle has these dynamics in both the x and y dimensions (Figure 2C).  In some simulations, the accuracy can break down due to rounding error when using smaller values.

The dynamical systems detected by an internal agent as described above match the mathematical description of the dynamics of physical trajectories or electrical RLC circuits. Thus, the linear dynamic matrices are not new. The main goal of the framework presented here is to describe how neural circuits could potentially use gating matrices to compute and represent the dynamics of spatiotemporal trajectories. This capacity of neural circuits would be important for behaviors such as catching a ball, or leaping onto a spinning platform. Note that a prediction matrix could be computed from three sequential positions of a single feature a(t), a(t+1) and a(t+2) and the next three positions, but in contrast to the framework presented above, a prediction matrix based on position alone will not generalize to arbitrary different locations of the trajectory.

This framework can model the dynamics of quadratic equations (with exponential solutions), but not higher level polynomials. However, higher level functions or multiple segments of different curves could be modeled by combining the above mechanisms with additional gating matrices to link segments of a trajectory that have different functional properties, such as the trajectory of a ball interacting with the floor or a wall. If the ball is sufficiently elastic it bounces when it strikes a hard surface such as the floor or a wall.  Thus, a collision with a wall will evoke different dynamics. These differential dynamics can be activated by the overlap of the object position with the position of the floor or a wall. The necessary components of the collision matrix can be computed from the sequential features on the objects, or  this function could use multiple simultaneous agents that predict different trajectory segments, and then generate a higher level gating matrix coding the relationships between different trajectory segments, or the initial vectors of different segments. This mechanism should also allow the coding of a curved line generated with Bezier matrices, with prediction and transition points when necessary defined by the length $k$ of individual segments.

In addition, trajectories could have branching points in which one trajectory splits into multiple trajectories. This could utilize branching matrices that are not square, but combine multiple square matrices that can split a single activity vector into multiple concatenated vectors that could be split into further vectors for branching. The splitting into different vectors could be used to describe the trajectory of a clod of dirt that breaks into pieces as it flies through the air or to describe the relationships between component elements of complex biological structures such as the branches of a tree.

**Edge following to find shape vertices.**

The example presented above makes the assumption that the indices and values for sequential elements of a trajectory can be detected accurately and automatically, allowing the computation of a prediction gating matrix between sequential elements without ambiguity.  Thus, the example presented in the previous section does not use a sampling process and does not directly address the value of the sensory input.

However, this cannot be used for generating a prediction gating matrix for features of a static image, where the sequential order between components of the image are not pre-defined. Instead, generating a prediction matrix about the features of a static image requires a sample gating matrix that samples the sensory input value at specific positions in an image in a consistent manner and matches them with an internal value (perhaps based on previous sample points). The generation of a consistent sampling or search pattern will allow the generation of a prediction matrix that can describe the sequentially sampled locations. In addition, the termination of the search can detect corners.

This section describes an example of the coding of figures for discovering higher order rules. The internal agent described here finds edges and follows edges to vertices. This could be used as a method to find sets of vertices that then could be linked at a higher level by gating matrices that can extrapolate to future shapes in the sequence, as described in the next section.

This uses the sample gating matrices described in the methods section. As shown in Equation 3, the current attentional focus location of the agent is defined by the vector $a$. The search or sampling strategy for the agent is defined by the activity pattern of sampling matrix $S$ and initial vector $s$. The matrix is raised the the power $m$ and multiplied by the initial vector, and then added to the current attentional focus vector to yield the sample test vector $o$. This generates a search or sampling pattern that moves progressively through the indexed space of the sensory map, allowing testing of the sensory value for different indexed locations.

For the example presented here, the sample gating matrix defines a sample test vector with a location moving in an expanding spiral. This uses the following internal sample gating matrix:

$$S = \begin{bmatrix} 1 & 1.5 & 0 & 0 \\ -v & 1-v & 0 & 0 \\ 0 & 0 & 1 & 1.5 \\ 0 & 0 & -v & 1-v \end{bmatrix}$$

where $v$ is the angular velocity (2*pi/32)^2 of the search pattern and 1.5 determines the rate of expansion of the spiral. Note that this matrix combines position and velocity in both the x and y dimension, in contrast to the first example, where position, velocity and acceleration are coded with separate matrices for the x and y dimensions (the computation of trajectories can be done either with a combined matrix or separate matrices). The multiplication $S^m s$ generates an expanding spiral over sampling time $m$ that starts from each current location $a$. The starting direction and subsequent direction of the expanding spiral depends upon the initial test location vector $s$. Different initial vectors allow the same sample gating matrix to expand in different initial directions and explore sample locations $o$ with clockwise or counterclockwise spin. The continuous dynamics of the sample gating matrix can also be discretized to the resolution of the pixels within an image to sample the sensory value of the image at that index.

The function of this search agent is shown in Figure 4. Here eight agents with different initial parameters are started within the dark gray shape (with grayscale value of 0.996). In this example, the internal agents are set to match the dark gray shade of the image. When the indices of the sample vector shift to another discrete pixel, the value of that pixel is matched with the internally stored range of the internal agent, and if it is within the image range of dark gray (e.g. 0.8 to 1.02), then the current index of the overall agent is updated to the new index. This means that the agents initially shift in their respective directions until they find an edge.

When each agent finds an edge, it proceeds along that edge dependent upon the internal dynamics of its equation. For example, a clockwise agent that initially moves in an upward direction will find a white pixel and then shift clockwise until it finds a dark gray pixel. This means that it will follow an edge in a rightward and upward direction. A counterclockwise agent that moves upward first will find an edge on the left and will move leftward and upward. In contrast, a clockwise agent that initially moves downward will move in a downward and leftward direction, and a counterclockwise agent that initially moves downward will move in a downward and righward direction.

Figure 4 illustrates the function of these agents on simple rectangular shapes. Figure 4A shows the trajectory of movement of 8 agents on a rectangle. They start in the center left and move in straight lines based on the initial direction of their vectors as they find gray pixels. Then when they reach the edges, they follow the edges in their characteristic direction, appearing as small, sharp, repeating loops until they encounter a vertex, where they will often not progress further due to their search pattern finding the same or adjacent pixels on each search or sampling cycle. Thus, these search agents are able to find and follow edges and find vertices and remain in vertices.  The input of sample and current focus vectors allows generation of a prediction matrix that can predict the direction of the edge being detected by individual agents, which usually match the orientation of one of the detected edges.  The detection of vertices in multiple shapes is shown by the circles in Figure 4B.  Most agents find a vertex and remain in a vertex in their steady state (though some get stuck on the edge away from a vertex). The detection of vertex locations can then be used as a framework for coding the relations between different groups of vertices (referred to as corners) in the form of gating matrices that code the components of the affine transformation.  This allows prediction of further components of the affine transformation such as size, translation and rotation, as described in the following section.

### Raven's task affine transformation

The framework of internal agents can be used to extrapolate the rules for transformations of shapes in the Raven's Progressive Matrices task.  In particular, the detection of vertices described in the previous section can provide the basis for encoding and prediction of a series of gating matrix relations for solving the Raven's task. (Note that to avoid confusion between the  transformation matrices used here and the position of a cognitive cue in the Raven's Progressive Matrices taks, the behavioral task is referred to below as the Raven's task.) Previous research has shown that a subset of problems in the Raven's task can be solved by applying affine transformations to the stimuli (Kunda et al., 2013). This section will focus on the learning and extrapolation of affine transformations to the task stimuli.

***Determination of relations between corner arrays.*** This partial solution of the affine components of the Raven's task involves the use of first level internal agents to follow the edges of shapes to find the vertices based on the sensory values, as described in the previous section of the results. (In this section, the sensory values are assumed to be the same for the individual shapes in black and white stimuli and are not considered further, but they could be a factor if the task uses color.) Then, higher level agents can detect the relations between the array of vertices (referred to as corners) in a single shape and the array of vertices (corners) between shapes. Detecting the relations between arrays of vertices requires that the vertices be grouped. The previous section identifies single column vectors for the location of single vertices, so the definition of a vertex is a single coordinate point. In contrast, this section defines a corner as a set of three column vectors defining three adjacent vertices on the shape.  If the corners are arranged in sequential order, the extraction of these relationships is more straightforward, but the simulations presented here assume the sampling of corners is random.  The next shape in a sequence can still be predicted even if corners are sampled randomly.

When learning the sequence of shapes in a Raven's progressive matrices task, the transition between shapes might involve an affine transformation such as a change in rotation or size. This can be represented as a transformation of the transformations of the corners.  In this framework, a second level internal agent may code the corners found by the first level agent, and then a third level internal agent codes the relationship between the representations in the second level internal agent.

Figure 5 shows an example of the representation of corners for this section. For example, one corner (with index c) of a rectangle might have the array matrix of vertex locations: $_{r=1}^{c=1}A = \begin{matrix} 2 & 4 & 4 \\ 4 & 4 & 1 \\ 1 & 1 & 1 \end{matrix}$ defined by the sensory index x and y of a different vertex in each column of the array matrix.   For coding of relations of corners,  each corner in each shape is coded as a transformation that starts from a baseline corner affine matrix made up of three location vectors (in three columns). This baseline corner is here defined separate from the individual shapes. This is illustrated in Figure 5, where the pyramid shaped corner is the baseline matrix $^{b}A = \begin{matrix} -1 & 0 & 1 \\ 0 & 1 & 0 \\ 1 & 1 & 1 \end{matrix}$.  The baseline corner is associated with the set of current corners in each shape defined by a location matrix. In this framework, the baseline corner can be linked to each current corner by a transform gating matrix r (at level r=2) computed from the inverse of the baseline array and the current corner array as $_{r=2}^{c}G = {}_{r=1}^{c}A(\,^{b}A^{-1})$.  In this single example, this yields the gating matrix at level 2 for corner c as $\substack{c=1\\r=2}G = \begin{matrix} 1 & 1 & 3 \\ -1.5 & 1.5 & 2.5 \\ 0 & 0 & 1 \end{matrix}$. Here the superscript c=1 indicates the specific corner and the preceding subscript r=2 indicates the level of the gate.

Different corners of the rectangle can then be related to each other by computing the next (r=3) level transformation matrix as $\substack{c=1,2\\r=3}G = \substack{c=2\\r=2}G(\substack{c=1\\r=2}G^{-1})$. For the second corner c=2 of a rectangle $\substack{c=2\\r=1}A = \begin{matrix} 4 & 4 & 2 \\ 4 & 1 & 1 \\ 1 & 1 & 1 \end{matrix}$ this gives the gating matrix as $\substack{c=2\\r=2}G = \begin{matrix} -1 & 1 & 3 \\ -1.5 & -1.5 & 2.5 \\ 0 & 0 & 1 \end{matrix}$. Using the two second level transformation matrices to compute the 3$^{rd}$ level transformation matrix gives $\substack{c=1\\r=3}G = \begin{matrix} 0 & -1 & 0 \\ 1 & 0 & 0 \\ 0 & 0 & 1 \end{matrix}$. Computing the rest of the rectangle shows that the 3$^{rd}$ level transformation matrix is the same for all stepwise interactions of corners in the rectangle. This means this rectangle can be defined by one corner, the baseline corner and the 3$^{rd}$ level transformation matrix. Similarly, other polygons can have similar relationships to their corners, though the matrices will vary for different corner configurations. For the example shown in Figure 4, different polygons can be defined by an initial level 2 gating matrix $\substack{c=2\\r=2}G = \begin{matrix} 1 & 0 & 0 \\ -\cos\left(\frac{2\pi}{s}\right) & 1 & 0 \\ 0 & 0 & 1 \end{matrix}$ and a 3$^{rd}$ level transformation matrix $\substack{c=1\\r=3}G = \begin{matrix} 0 & -1 & 0 \\ 1 & 2\cos\left(\frac{2\pi}{s}\right) & 0 \\ 0 & 2\cos\left(\frac{2\pi}{s}\right) & 1 \end{matrix}$, where s is number of sides.

This example works for creating polygons. However, as the number of sides increases there is stretching in different dimensions, so this is not the formula for perfectly symmetrical polygons.

***Extrapolation of affine transformations for Raven's task.*** In the simulation presented here, the extrapolation of shapes is accomplished by random selection of different corners. In the simulation shown in Figure 6, the selection process starts with corner 1 as $^{b=1}A$, and then shifts to a different random corner with a random index c selected between the first and last corner. The simulation then computes the gating matrix between these two corners as $\substack{c=1\\r=2}G = {}^{c=2}A\,{}^{b=1}A^{-1}$. Subsequently, the network creates $^{b=2}A$ using the previous c vector $^{c=2}A$ and selects a new random index c for $^{c=3}A$. Once it computes $^{c=2}_{r=2}G = {^{c=3}A}(^{b=2}A)^{-1}$ for this new pair of corners, it can then use the two second level gating matrices described in the previous two sentences to compute the gating transformation matrix at level 3 as $^{c=1,2}_{r=3}G = {^{c=2}_{r=2}G}(^{c=1}_{r=2}G)^{-1}$. This third level transformation between randomly selected corners is then stored in an episodic buffer. The process then repeats for subsequent steps, so that after a period of time, the episodic buffer has a sequence of transformation elements that essentially form relational transformations jumping randomly between different stored elements of the sequence.

After this process has been completed for a number of steps, the buffer can be used to recreate the sequence of relational corner plotting from any arbitrary position on the array of corners. If the starting position of the recreation is toward the end of the sequence, then the buffer extrapolates the affine transformation to predict the shape, size and rotation of shapes beyond the end of the original set of shapes, as shown by the gray lines in Figure 6. Thus, a sequence of third level gating transformations can be used to predict the next step in an affine transformation for the Raven's task. The selection of a correct response in the Raven's task would then require an additional matching operation to relate the extrapolated shape to the correct matrix response in the Raven's task, which usually provides an array of possible answers.

**DISCUSSION**

The framework proposed here focuses on using activity patterns of internal neural agents to explore and fit functions to the structure of sensory input and the internal structure of other internal gating matrices. This provides the flexibility for a network to have many internal agents simultaneously fitting multiple features of a sensory input at different hierarchical levels, and the capacity to extend these fitting functions to the next step in a sequence or matrix, as tested in the example of affine transformations such as rotation, size and translation that are relevant to performance of the Raven's progressive matrices task.

The solutions to affine transformations that are used in the Raven's progressive matrices task shown here have an affine matrix structure similar in qualitative function to the affine model used to solve this task (Kunda et al., 2013). However, in contrast to that previous work, the current paper focuses on generating these affine models using internal neural agents that can learn to form the rules for affine transformations. Many aspects of the Raven's task have not yet been addressed, such as number and set, but the same framework could be used for that.  For example, the completion of set stimuli could be obtained by using repeating circular boundary conditions to solve problems. For example, a row containing circle, square, triangle, followed by a row with square, triangle, circle could be result in completion of a row with only triangle, circle.  The completion of textures within different shapes could be generated by internal agents that extrapolate repeated elements in two directions. Other models have addressed the Raven's progressive matrices based on direct processing of the visual images (Lovett et al., 2009; Raudies and Hasselmo, 2017; Barrett et al., 2018) using techniques such as automated sketch understanding (Lovett et al., 2009) or training of relation matrices (Santoro et al., 2017; Barrett et al., 2018). In contrast to these approaches, another set of models have used pre-coding of the visual transformations present in the Raven's stimuli to form models of the solution of the task (Carpenter et al., 1990; Rasmussen and Eliasmith, 2011).

On a neural level, the proposed interaction of higher level matrix outputs with lower level matrix elements resembles the multiplicative interactions of synapses in the extensive biophysical and computational modelling of the interaction of synapses by Bartlett Mel and colleagues (Mel, 1993; Poirazi et al., 2003).  In those previous models, the multiplicative interaction of synapses was due to the voltage-dependent properties of NMDA receptors. In those models, NMDA channels at synapse A would not be activated by presynaptic input A alone, but required the post-synaptic depolarization caused by an adjacent synapse B to remove magnesium block at synapse A, thus corresponding to a multiplicative interaction of the presynaptic input at synapse A with the input at synapse B. The use of multiplicative units in earlier neural networks was referred to as product units (Durbin and Rumelhart, 1989) or sigma-pi networks (Courrieu, 2004).  A multiplication function can also be generated at a network level by interactions of nonlinear input-output functions with feedback inhibition (Nezis and van Rossum, 2011). A multiplicative gating of interactions could also be mediated by the interaction between neural populations with different patterns of rhythmic activity, such as gamma or beta frequency oscillations (Buschman et al., 2012;

Lundqvist et al., 2016; Lundqvist et al., 2018; Sherfey et al., 2018). Functional magnetic resonance imaging in humans performing variants of the Raven's task have shown neural activity in cortical regions including prefrontal and parietal cortex as well as basal ganglia that could correspond to the activity-based gating matrices described here (Melrose et al., 2007). The gating mechanisms and hierarchical interaction of gating matrices used here was inspired by previous models of specific cognitive tasks (Badre et al., 2010; Badre and Frank, 2012; Chang et al., 2017; Bhandari and Badre, 2018).

This framework requires the computation of the inverse of a matrix in local neural circuits. Recurrent neural networks for computation of the inverse of a matrix have been designed (Wang, 1993; Wang, 1997). Dynamical systems for other complex matrix operations have been developed that could be relevant to the function of these circuits (Brockett, 1991). The example of trajectory prediction coding resembles autoregressive models. The properties of internal agents propagating as activity that generates other patterns of activity is similar to the activity patterns in Conway's Game of Life (Reia and Kinouchi, 2014), in which patterns of activity can create other patterns of activity (such as the glider gun).

As noted in the introduction, these models could be related to the previous models of episodic memory. In those models, individual locations along a spatiotemporal trajectory were associated with specific objects or events coded on these trajectories (Hasselmo, 2009; Hasselmo et al., 2010; Hasselmo, 2012). This forms a framework that can be extended to the coding of semantic representations of the environment by the relations between different component features. In these previous models of episodic memory, the spatial location of an object or event was coded by a grid cell model developed by Burgess and colleagues that used the relative phases of oscillations (Burgess et al., 2007), or the spiking activity relative to oscillation phase (Hasselmo, 2008). The coding of time in episodic memory can use time cells firing at specific temporal intervals (Howard et al., 2014; Liu et al., 2018; Mau et al., 2018). The framework presented here builds on that earlier model of episodic memory (Hasselmo, 2009; Hasselmo et al., 2010; Hasselmo, 2012), and extends it to a more general framework for the semantic relationship between elements of sensory input as suggested in a previous paper as follows (Hasselmo et al., 2010): "The example presented here focused on the linear coding of spatial dimensions, but the same properties could be applied to other dimensions. For example, a change in brightness could drive firing frequency to cause a phase shift coding the state of brightness, or a change in color could drive a phase shift to code the state of color. Similarly, cells responding to angular velocity could drive a phase shift that codes the shift in visual angle of objects in the visual field. Even complex actions such as expansion could be coded by a population of neurons. The population coding expansion would drive a change in firing frequency in neurons coding the width of an object such as a balloon, causing a progressive change in relative phase coding the change in width of the balloon" (Hasselmo et al., 2010).

The use of a location code in the sensory map for coding of shapes is related to the proposal that location of individual features is essential to coding an object (Hawkins et al., 2017). For example, when recognizing a coffee cup, the position of fingers detecting surfaces at specific allocentric locations is important. That model codes objects by the allocentric position of each feature, forming associations between all of these allocentric positions such that sensing one corner of a cube will predict the sensation of other corners of the cube.

Note that the model presented here focuses on exploring a network model that uses rapid gating by patterns of activity rather than synaptic modification based on gradient descent learning. This differs from procedures used in deep learning to explore new domains of model space. Reviews of network modelling research focus on the need for a causal physical model of the world that allows flexible adaptation to different task demands and the use compositionality and learning-to-learn (Lake et al., 2017), and the need to make inferences and extrapolation based on small data sets (Marcus, 2018). There are many previous models that use elements of activity dependent gating in network design. For example, LSTM circuits have the capacity to allow flexible gating of information into working memory (Hochreiter and Schmidhuber, 1997; Schmidhuber, 2015).

The generation of internal functions by internal agents resembles certain aspects of computer graphics and computer vision. For computer graphics, the generation of design curves can use Bezier curves, which can be implemented with a neural type matrix representation (Joy, 2000; Buse, 2013). This has been used to capture outlines of visual images (Pal et al., 2004; Safraz and Masood, 2007). Visual surfaces can also be modelled using partial differential equations (Ugail et al., 1999), or Non-uniform Rational B-splines (Schumaker, 1981), which can be extracted from the images algorithmically (Ravari and Taghirad, 2016). The process of understanding an image as a set of functions can also be related to vectorization of raster/pixel images as in the program potrace (Selinger, 2003).

Further development of this model will address the specific biophysical properties of neural mechanisms for gating and routing that are not yet modeled in detail. In addition, further development will address the generation of complex internal agents evaluated on the basis of their capacity to predict increasing ranges of features on multiple levels and scales in sensory input.

### Figure legends

**Figure 1.** Diagram of the properties of internal agents.  The sensory space at bottom shows a rectangle image stimulus. As shown, an internal agent can include both a prediction gating matrix and a sample matrix. The prediction gating matrix takes input from sensory input arrays X that combine column vectors representing individual locations and sensory values. The prediction matrix takes input from two sequential focus positions and codes relations between these arrays. The search matrix determines the next location being sampled from the sensory map and compares with a stored value, and can then provide this as the next focus location. The prediction gating matrix can be raised to powers to predict multiple future locations of that sensory value. The higher level internal agents perform the same type of operation on the values of elements of the lower level prediction matrices.

**Figure 2.** Trajectories that can be effectively predicted with gating matrices. The open circles show trajectories generated by different functions to be predicted. The asterisks within the circles show the successful prediction of the trajectories using the gating matrices described here. A. Parabolic trajectory. B. sine wave trajectory, C. circular trajectory, D. cosine function plotted relative to x squared, E. sinewave multiplied by an exponential, F. inward spiral, G. straight line, H. cosine function in x dimension plotted.

The trajectories are all predicted starting from position input and using internal gating matrices to compute velocity and acceleration and to compute the dynamic matrices predicting the variables for that dimension.

**Figure 3.** Schematic of internal agent gating matrices that can generate the prediction of trajectories. A. Summary of components of a single internal agent operating on multiple levels. Note that each level involves a computation from the level below, and a buffer holding previous computations for use in the next higher level. From bottom up, the agent receives an array of vectors X that code three feature locations on the moving object. The locations at consecutive time points are used to compute the second level gating matrix $_2G$, which obtains velocity in the third column. Consecutive second level matrices $_2G$ containing velocity are combined to compute a third level matrix $_3G$ which includes acceleration in the third column. The position, velocity and acceleration are then transformed into a third level array $_3X$ used to compute the fourth level prediction matrix $_4G$.  B. The sequential computations by the agent at each time step are illustrated. At the highest level, the dynamic prediction matrix can then generate the remainder of the trajectory.

**Figure 4**. Examples of internal search agents following the edges and finding vertices of shapes. A. Eight agents are initiated in the middle left of a rectangle shape. They move in eight different directions based on their initial x and y direction and their clockwise or counterclockwise search trajectory. The agents are coded to update their index when they find the dark gray shade (between 0.8 and 1.02), so they move in a straight line across the dark gray pixels until they hit an edge. When they hit an edge, the direction of their search determines the direction of their edge following.  B. The circles indicate the locations where agents stabilize on the shape after extended simulation.  Most agents stop moving their location when they reach a vertex, due to their search pattern bringing them back to the same pixel or an adjacent pixel.

**Figure 5.** Example of shape corners coded relative to a baseline corner. Each corner is defined by the coordinates of three sequential vertices of the rectangle.  The example shows three corners of the rectangle

in upper right with thicker black lines and larger squares for the first corner, and the thinnest line and smallest squares for the third corner. The baseline [-1 0 1; 0 1 1; 1 0 1] is shown in lower left with solid black lines.

**Figure 6**. Examples of shapes undergoing affine transformations.  In each row, a sample shape shown with solid black lines undergoes two steps of translation and additional affine transformations. The dashed gray lines show the model of the shape generated by internal agents, and the prediction of subsequent shapes generated by the gating matrices that link current corners of the shape with potential future corners of the shape.  The very thin gray lines show the sequential relations coded by the gating matrices for each corner relation.  A. Triangle showing two steps of translation, rotation and size increase (solid black lines) and showing correct prediction of future rotation and size change (dashed gray lines).  B. Rectangle showing two steps of translation, rotation and size increase (solid black) and showing correct prediction of future rotation and size change (dashed gray).  C. Rectangle showing two steps of translation and size increase without rotation, and correction prediction of size increase.  D. Pentagon showing two steps of translation and rotation (solid black lines), and distorted prediction of shape and future rotation (dashed gray lines).  Note that due to inaccurate prediction, the dashed prediction lines do not perfectly line up with the shape.

**Bibliography**

Badre D, Frank MJ (2012) Mechanisms of hierarchical reinforcement learning in cortico-striatal circuits 2: evidence from fMRI. Cereb Cortex 22:527-536.
Badre D, Kayser AS, D'Esposito M (2010) Frontal cortex and the discovery of abstract action rules. Neuron 66:315-326.
Barrett DT, Hill F, Santoro A, Morcos AS, Lillicrap T (2018) Measuring abstract reasoning in neural networks. In: Proceedings of the 35th International Conference on Machine Learning. Stockholm, Sweden.
Bhandari A, Badre D (2018) Learning and transfer of working memory gating policies. Cognition 172:89-100.
Brockett RW (1991) Dynamical systems that sort lists, diagonalize matrices, and solve linear programming problems. Linear Algebra Appl 146:79-91.
Burgess N, Barry C, O'Keefe J (2007) An oscillatory interference model of grid cell firing. Hippocampus 17:801-812.

Buschman TJ, Denovellis EL, Diogo C, Bullock D, Miller EK (2012) Synchronous oscillatory neural ensembles for rules in the prefrontal cortex. Neuron 76:838-846.

Buse L (2013) Implicit matrix representations of rational Bezier curves and surfaces. In: SIAM Conference on Geometric and Physical Modeling, pp 14-24.

Bush D, Barry C, Manson D, Burgess N (2015) Using Grid Cells for Navigation. Neuron 87:507-520.

Byrne P, Becker S, Burgess N (2007) Remembering the past and imagining the future: a neural model of spatial memory and imagery. Psychol Rev 114:340-375.

Carpenter PA, Just MA, Shell P (1990) What one intelligence test measures: a theoretical account of the processing in the Raven Progressive Matrices Test. Psychol Rev 97:404-431.

Chang AE, Ren Y, Whiteman AS, Stern CE (2017) Evaluating the relationship between activity in frontostriatal regions and successful and unsuccessful performance of a contextdependent rule learning task. Society for Neuroscience Abstracts 43.

Chen TW, Wardill TJ, Sun Y, Pulver SR, Renninger SL, Baohan A, Schreiter ER, Kerr RA, Orger MB, Jayaraman V, Looger LL, Svoboda K, Kim DS (2013) Ultrasensitive fluorescent proteins for imaging neuronal activity. Nature 499:295-300.

Chen X, Leischner U, Rochefort NL, Nelken I, Konnerth A (2011) Functional mapping of single spines in cortical neurons in vivo. Nature 475:501-505.

Courrieu P (2004) Solving time of least square systems in sigma-pi unit networks. Neural Information Processing - Letters and Reviews 4:39-45.

Cutsuridis V, Hasselmo M (2012) GABAergic contributions to gating, timing, and phase precession of hippocampal neuronal activity during theta oscillations. Hippocampus 22:1597-1621.

Douglas RJ, Martin KA, Witteridge D (1989) A canonical microcircuit for neocortex. Neural Comput 1:480-488.

Durbin R, Rumelhart DE (1989) Product units: A computational powerful and biologically plausible extension to backpropagation networks. Neural Computation 1:133-142.

Eliasmith C (2013) How to Build a Brain: A Neural Architecture for Biological Cognition. New York, NY: Oxford University Press.

Eliasmith C, Stewart TC, Choo X, Bekolay T, DeWolf T, Tang Y, Rasmussen D (2012) A large-scale model of the functioning brain. Science 338:1202-1205.

Erdem UM, Hasselmo M (2012) A goal-directed spatial navigation model using forward trajectory planning based on grid cells. Eur J Neurosci 35:916-931.

Erdem UM, Hasselmo ME (2014) A biologically inspired hierarchical goal directed navigation model. J Physiol Paris 108:28-37.

Erdem UM, Milford MJ, Hasselmo ME (2015) A hierarchical model of goal directed navigation selects trajectories in a visual environment. Neurobiol Learn Mem 117:109-121.

Fodor JA, Pylyshyn ZW (1988) Connectionism and cognitive architecture: a critical analysis. Cognition 28:3-71.

Hasselmo ME (2005) A model of prefrontal cortical mechanisms for goal-directed behavior. J Cogn Neurosci 17:1115-1129.

Hasselmo ME (2008) Grid cell mechanisms and function: contributions of entorhinal persistent spiking and phase resetting. Hippocampus 18:1213-1229.

Hasselmo ME (2009) A model of episodic memory: mental time travel along encoded trajectories using grid cells. Neurobiol Learn Mem 92:559-573.

Hasselmo ME (2012) How we remember: Brain mechanisms of episodic memory. Cambridge, MA: MIT Press.

Hasselmo ME, Eichenbaum H (2005) Hippocampal mechanisms for the context-dependent retrieval of episodes. Neural Networks 15:689-707.

Hasselmo ME, Stern CE (2018) A network model of behavioural performance in a rule learning task. Philos Trans R Soc Lond B Biol Sci 373.

Hasselmo ME, Giocomo LM, Brandon MP, Yoshida M (2010) Cellular dynamical mechanisms for encoding the time and place of events along spatiotemporal trajectories in episodic memory. Behav Brain Res 215:261-274.

Hawkins J, Ahmad S, Cui Y (2017) A Theory of How Columns in the Neocortex Enable Learning the Structure of the World. Front Neural Circuits 11:81.

Hochreiter S, Schmidhuber J (1997) Long short-term memory. Neural Comput 9:1735-1780.

Howard MW, MacDonald CJ, Tiganj Z, Shankar KH, Du Q, Hasselmo ME, Eichenbaum H (2014) A unified mathematical framework for coding time, space, and sequences in the hippocampal region. J Neurosci 34:4692-4707.

Hummel JE, Biederman I (1992) Dynamic binding in a neural network for shape recognition. Psychol Rev 99:480-517.

Hummel JE, Holyoak KJ (1997) Distributed representations of structure: a theory of analogical access and mapping. Psychol Rev 104:427-466.

Joy KI (2000) A matrix formulation of the cubic Bezier curve. In: On-line geometric modeling notes.

Klausberger T, Magill PJ, Marton LF, Roberts JD, Cobden PM, Buzsaki G, Somogyi P (2003) Brain-state- and cell-type-specific firing of hippocampal interneurons in vivo. Nature 421:844-848.

Koene RA, Hasselmo ME (2005) An integrate-and-fire model of prefrontal cortex neuronal activity during performance of goal-directed decision making. Cereb Cortex 15:1964-1981.

Kriete T, Noelle DC, Cohen JD, O'Reilly RC (2013) Indirection and symbol-like processing in the prefrontal cortex and basal ganglia. Proc Natl Acad Sci U S A 110:16390-16395.

Kubie JL, Fenton AA (2012) Linear look-ahead in conjunctive cells: an entorhinal mechanism for vector-based navigation. Front Neural Circuits 6:20.

Kunda M, McGreggor K, Goel AK (2013) A computational model for solving problems from the Raven's Progressive Matrices intelligence test using iconic visual representations. Cognitive Systems Research 22-23:47-66.

Lake BM, Ullman TD, Tenenbaum JB, Gershman SJ (2017) Building machines that learn and think like people. Behavioral and Brain Sciences 40:e253.

LeCun Y, Bengio Y, Hinton G (2015) Deep learning. Nature 521:436-444.

Liu Y, Tiganj Z, Hasselmo ME, Howard MW (2018) A neural microcircuit model for a scalable scale-invariant representation of time. Hippocampus in press.

Lovett A, Tomai E, Forbus K, Usher J (2009) Solving geometric analogy problems through two-stage analogical mapping. Cogn Sci 33:1192-1231.

Lundqvist M, Herman P, Warden MR, Brincat SL, Miller EK (2018) Gamma and beta bursts during working memory readout suggest roles in its volitional control. Nature communications 9:394.

Lundqvist M, Rose J, Herman P, Brincat SL, Buschman TJ, Miller EK (2016) Gamma and Beta Bursts Underlie Working Memory. Neuron 90:152-164.

Marcus G (2018) Deep learning: A critical appraisal. In. ArXiv.

Mau W, Sullivan DW, Kinsky NR, Hasselmo ME, Howard MW, Eichenbaum H (2018) The Same Hippocampal CA1 Population Simultaneously Codes Temporal Information over Multiple Timescales. Curr Biol 28:1499-1508 e1494.

Mel BW (1993) Synaptic integration in an excitable dendritic tree. J Neurophysiol 70:1086-1101.

Melrose RJ, Poulin RM, Stern CE (2007) An fMRI investigation of the role of the basal ganglia in reasoning. Brain Res 1142:146-158.

Miller EK, Cohen JD (2001) An integrative theory of prefrontal cortex function. Annual Review of Neuroscience 24:167-202.

Mountcastle VB (1997) The columnar organization of the neocortex. Brain 120 ( Pt 4):701-722.

Nezis P, van Rossum MC (2011) Accurate multiplication with noisy spiking neurons. J Neural Eng 8:034005.

O'Reilly RC, Frank MJ (2006) Making working memory work: a computational model of learning in the prefrontal cortex and basal ganglia. Neural Comput 18:283-328.

Pal S, Biswas PK, Abraham A (2004) Face recognition using interpolated Bezier curve based representation. In: Proceedings of the International Conference on Information Technology: Coding and Computing (ITCC'04).

Poirazi P, Brannon T, Mel BW (2003) Arithmetic of subthreshold synaptic summation in a model CA1 pyramidal cell. Neuron 37:977-987.

Rasmussen D, Eliasmith C (2011) A neural model of rule generation in inductive reasoning. Topics in cognitive science 3:140-153.

Raudies F, Hasselmo ME (2015) Differences in Visual-Spatial Input May Underlie Different Compression Properties of Firing Fields for Grid Cell Modules in Medial Entorhinal Cortex. PLoS Comput Biol 11:e1004596.

Raudies F, Hasselmo ME (2017) A model of symbolic processing in Raven's progessive matrices. Biologically Inspired Cognitive Architectures 21:47-58.

Raudies F, Hinman JR, Hasselmo ME (2016) Modelling effects on grid cells of sensory input during self-motion. J Physiol 594:6513-6526.

Ravari RN, Taghirad HD (2016) NURBS-based representation of urban environments for mobile robots. In: 4th International Conference on Robotics and Mechatronics (ICROM), pp 20-25.

Reia SM, Kinouchi O (2014) Conway's Game of Life is a near-critical metastable state in the multiverse of cellular automata. Phys Rev E Stat Nonlin Soft Matter Phys 89:052123.

Rockland KS (2010) Five points on columns. Front Neuroanat 4:22.

Safraz M, Masood A (2007) Capturing outlines of planar images using Bezier cubics. Computers and Graphics 31:719-729.

Santoro A, Raposo D, Barrett DG, Malinoswki M, Pascanu R, Battaglia P, Lillicrap T (2017) A simple neural network module for relational reasoning. In: Neural Information Processing Systems. Long Beach, CA.

Schmidhuber J (2015) Deep learning in neural networks: an overview. Neural Netw 61:85-117.

Schumaker LL (1981) Spline Functions: Basic Theory. New York, NY: John Wiley and Sons.

Selinger P (2003) Potrace: a polygon-based tracing algorithm. In.

Sherfey JS, Ardid S, Hass J, Hasselmo ME, Kopell NJ (2018) Flexible resonance in prefrontal networks with strong feedback inhibition. PLoS Comput Biol 14:e1006357.

Siegel M, Warden MR, Miller EK (2009) Phase-dependent neuronal coding of objects in short-term memory. Proc Natl Acad Sci U S A.

Ugail H, Bloor MIG, Wilson MJ (1999) Techniques for interactive design using the PDE method. In: ACM Transactions on Graphics, pp 195-212.

van der Velde F, de Kamps M (2006) Neural blackboard architectures of combinatorial structures in cognition. Behavioral and Brain Sciences 29:37-70; discussion 70-108.

van der Velde F, de Kamps M (2015) The necessity of connection structures in neural models of variable binding. Cogn Neurodyn 9:359-370.

von der Malsburg C (1995) Binding in models of perception and brain function. Curr Opin Neurobiol 5:520-526.

Wallis JD, Miller EK (2003) From rule to response: neuronal processes in the premotor and prefrontal cortex. J Neurophysiol 90:1790-1806.

Wallis JD, Anderson KC, Miller EK (2001) Single neurons in prefrontal cortex encode abstract rules. Nature 411:953-956.

Wang J (1993) Recurrenet neural networks for solving linear matrix equations. Computers Math Applic 26:23-34.

Wang J (1997) Recurrent neural networks for computing pseudoinverses of rank-deficient matrices. SIAM J Sci Comput 18:1479-1493.

Zilli EA, Hasselmo ME (2008) Modeling the role of working memory and episodic memory in behavioral tasks. Hippocampus 18:193-209.

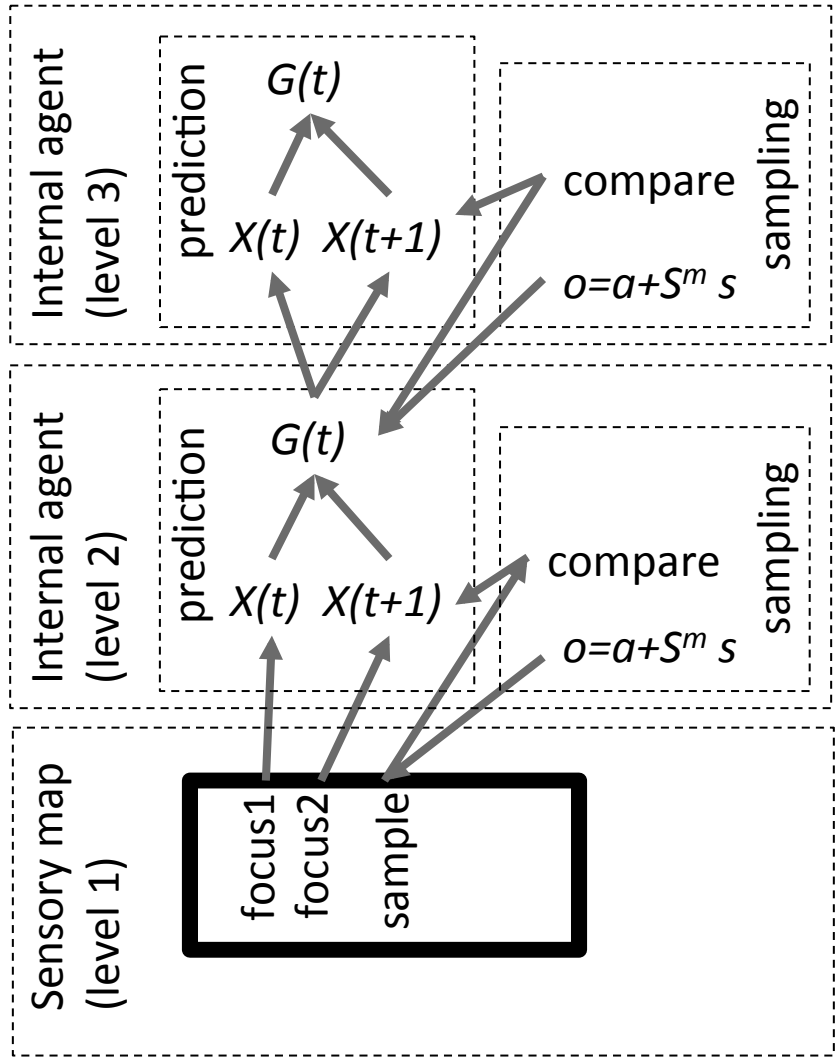

Figure 1

A

B

C

D

E

F

G

H

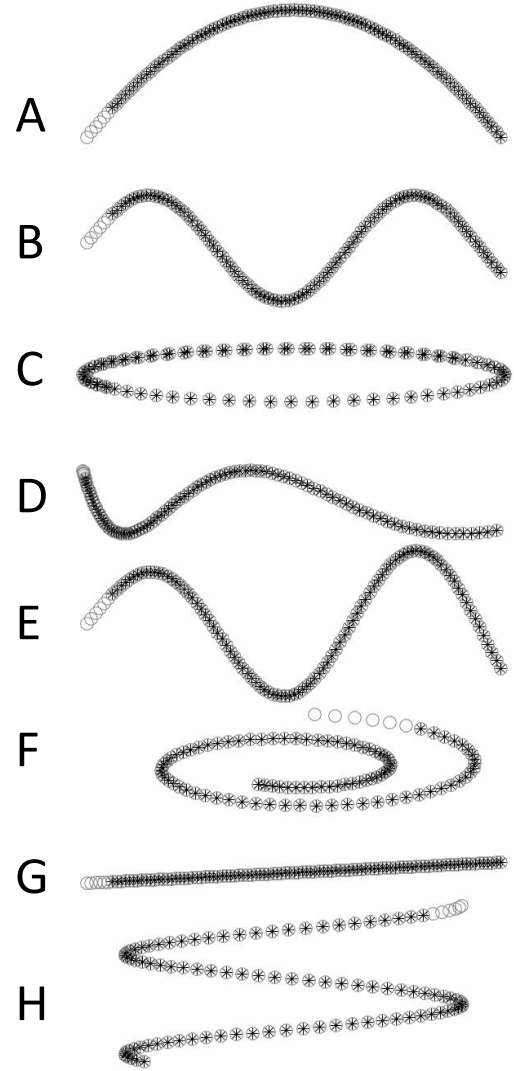

Figure 2

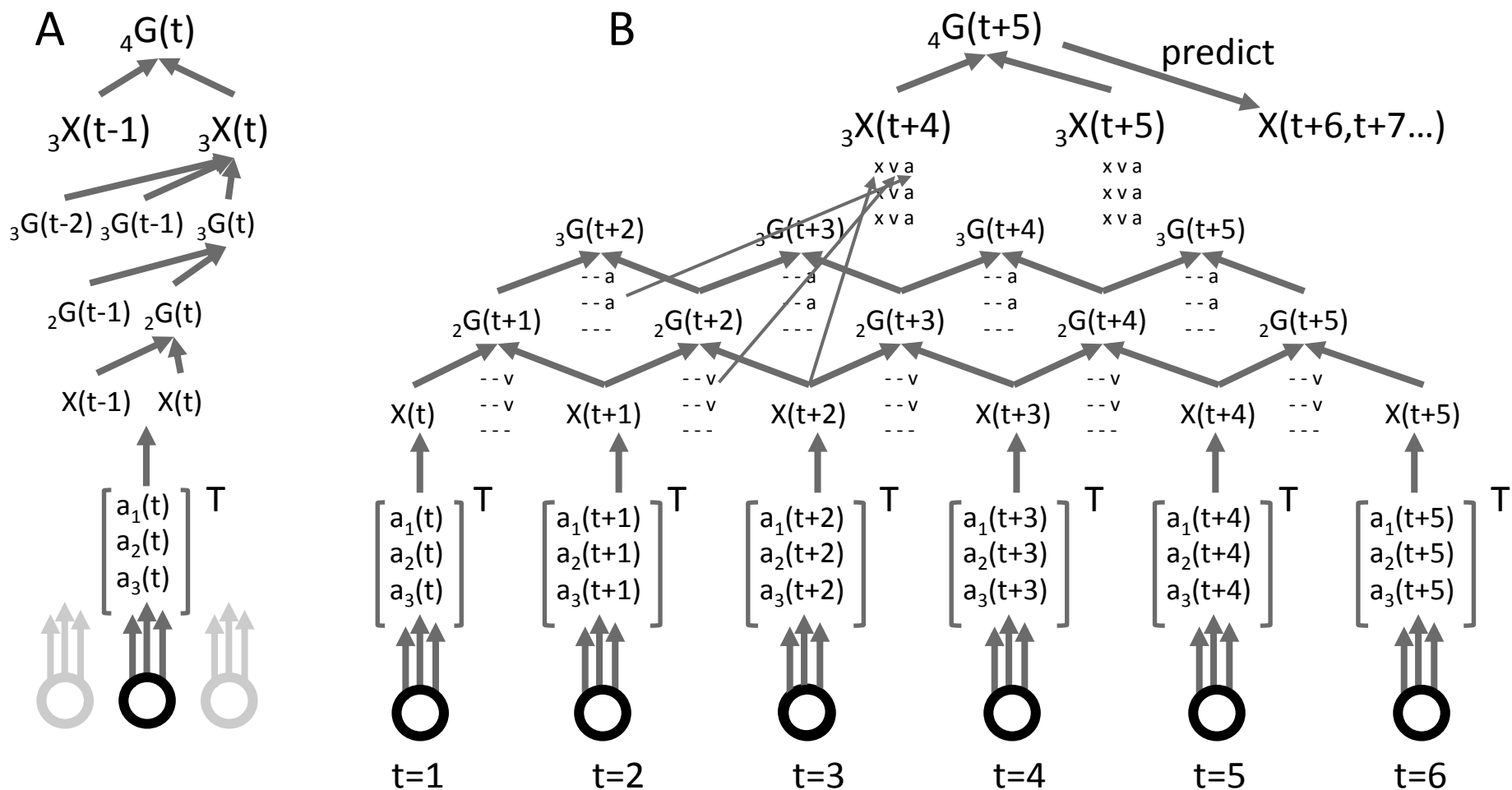

Figure 3

A

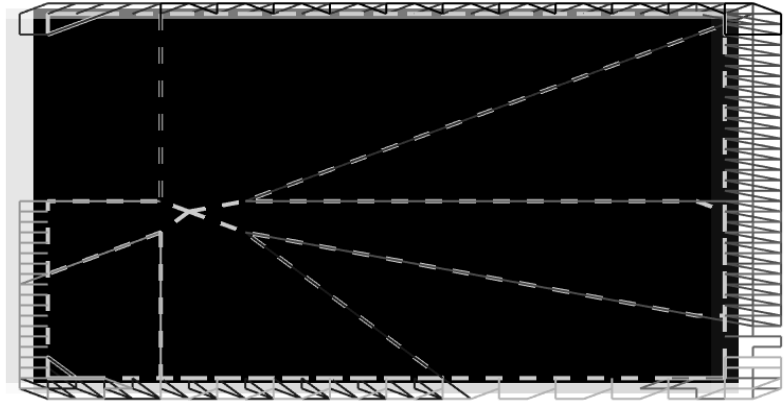

B

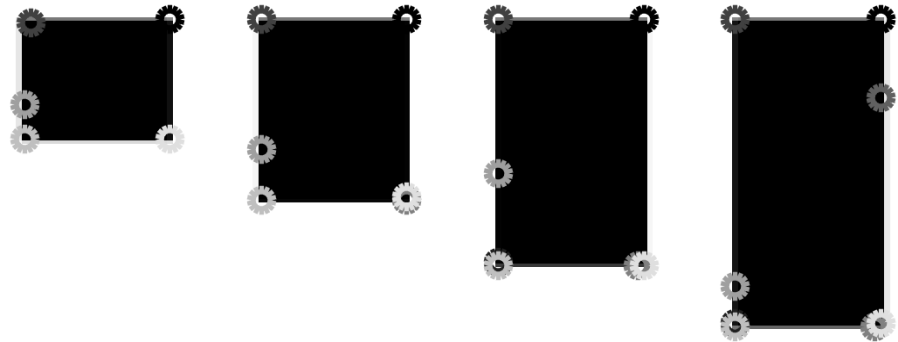

Figure 4

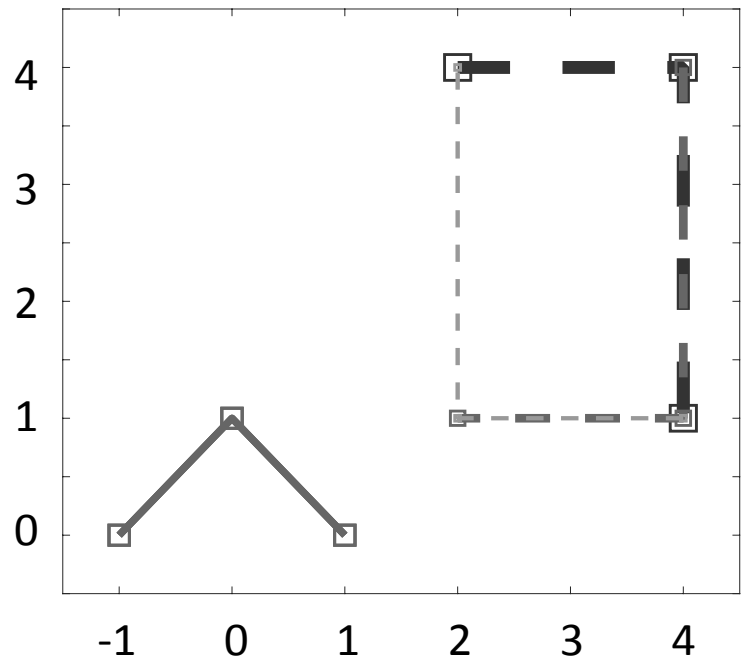

Figure 5

A 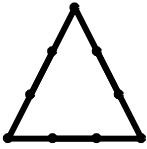 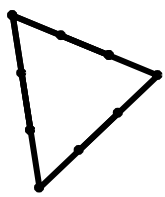 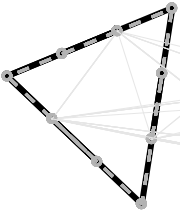 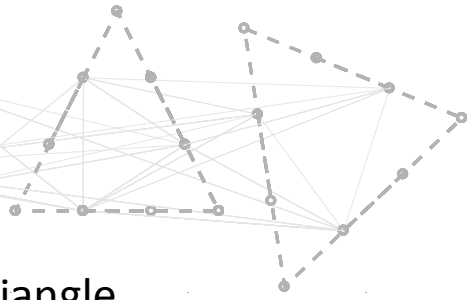

Size, rotation of triangle

B 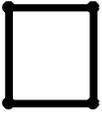 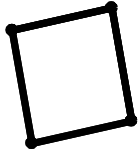 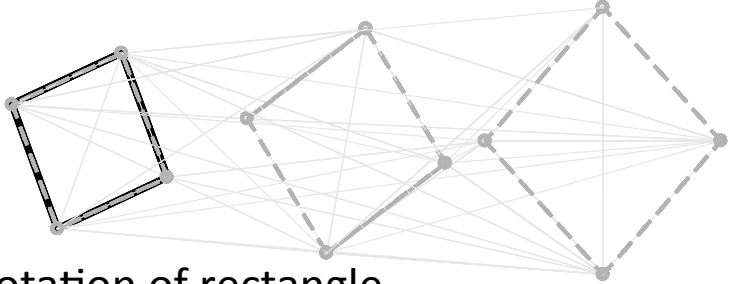

Size and rotation of rectangle

C 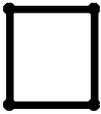 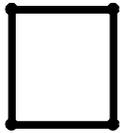 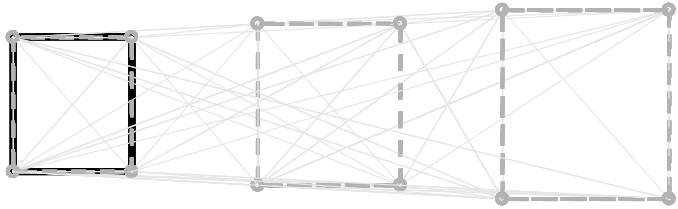

Size of rectangle

D 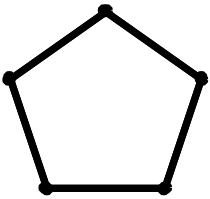 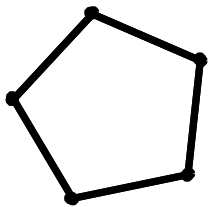 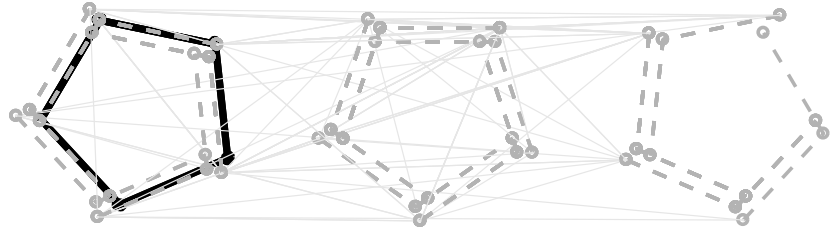

Rotation of pentagon

Figure 6